\documentclass[
  ,final            % use final for the camera ready runs
  ,numberedheadings % uncomment this option for numbered sections
  ]
  {aipproc}

\layoutstyle{6x9}

\usepackage{amsmath, amssymb}
\usepackage{pifont}

\begin{document}

\title{Activity and rotation of low mass stars in young open clusters}


\author{Andreas Seifahrt}{
 address={Institut f\"ur Astrophysik, Georg-August-Universit\"at, D-37077 G\"ottingen, Germany}
}

\author{Ansgar Reiners}{
address={Institut f\"ur Astrophysik, Georg-August-Universit\"at, D-37077 G\"ottingen, Germany}
}

\author{Aleks Scholz}{
address={SUPA, School of Phys. and Astron., Univ. of St Andrews, North Haugh, St Andrews,  KY16 9SS, UK}
}

\author{Gibor Basri}{
address={Astronomy Department, University of California, Berkeley, CA 94720}
}

\begin{abstract}
We present first results from a multi-object
spectroscopy campaign in IC2602, the Hyades, the
Pleiades, and the Coma cluster using VLT/FLAMES. We analysed the data for radial velocity,
rotational velocity ($v\sin{i}$), and H$\alpha$-activity. Here, we highlight three aspects 
of this study in the context of rotational braking and the rotation-activity
relationship among low mass stars. Finally we discuss the cluster membership of sources in IC2602.
\end{abstract}

\maketitle

%%%%%%%%%%%%%%%%%%%%%%%%%%%%%%%%%%%%%%%%%%%%
%% MAINMATTER
%%%%%%%%%%%%%%%%%%%%%%%%%%%%%%%%%%%%%%%%%%%%

\section{Rotational braking in the Hyades}

Stars gain angular momentum by hydrostatic contraction
during their formation phase. Thus, young stars are
mostly fast rotators.
Magnetic fields drive stellar winds that lead to a
rotational braking and a spin-down
of stars on the main
sequence, following the empirically established
``Skumanich law'' ($\omega\sim t^{-0.5}$, \cite{1972ApJ...171..565S}). This
picture, however, breaks down at very low masses. Old
field stars later than M4 are often fast rotators (e.g.,\cite{1998A&A...331..581D,2003ApJ...583..451M}). 
Rotational braking apparently is ineffective at masses $<0.3$\,M$_{\odot}$, 
indicating a change in the wind properties and/or the 
magnetic field. This mass threshold between slow and rapid rotation is
likely age dependent and seems to be shifted to earlier
spectral types (higher masses) for stars of intermediate
age between the ZAMS and about 1\,Gyr (e.g.,\cite{1999ASPC..158...63H}). 
We have found a Hyades member (VA 486, spectral
type M1V) to be a rapid rotator ($v\sin{i}\sim22$\,km/s) from a
UVES spectra obtained during our FLAMES campaign.
This star, and the previously known fast rotating K8V
Hyades member VB 190 \cite{1987ApJ...321..459R}, indicate
that at the age of the Hyades ($\sim550$\,Myrs) the threshold
of efficient magnetic braking seems to be shifted to
$\sim0.6$\,M$_{\odot}$ (see Fig 1).

\begin{figure}[ht]
\resizebox{0.8\textwidth}{!}{
\includegraphics[]{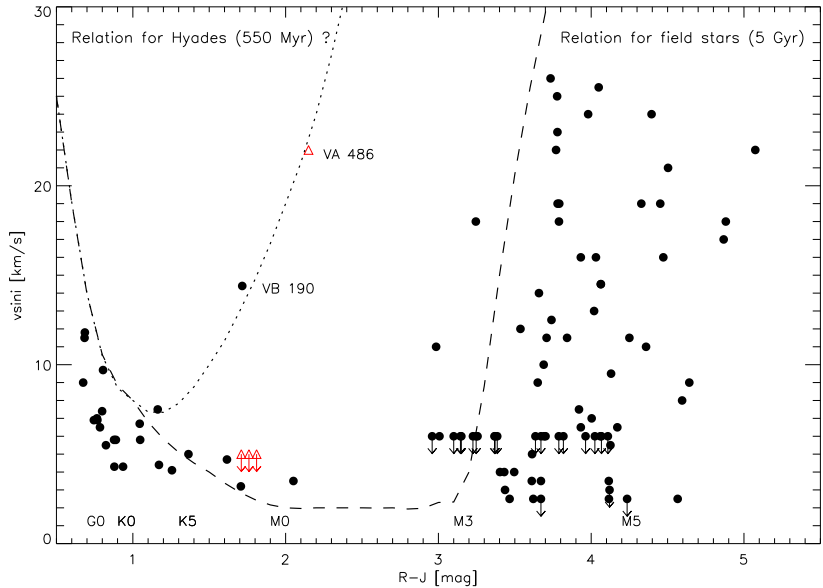}
}
\caption{Rotational velocity ($v\sin{i}$) for Hyades members. Data for
stars earlier than M0 are from Radick et al. \cite{1987ApJ...321..459R}, data for stars
later than M0 from Reid \& Mahoney \cite{2000MNRAS.316..827R}. The red triangles
denote the $v\sin{i}$ measurements obtained from our FLAMES
sample. The existence of fast rotators with spectral types earlier
than M3V indicate a change in the efficiency of magnetic braking
at the age of the Hyades ($\sim550$\,Myrs, dotted line) when compared
to the old field population ($\sim$5\,Gyrs, dashed line). More data
points for stars between M0V--M5V are needed to back up this
finding.}
\end{figure}

\section{Rotation and activity in the Coma cluster}

Another aspect of the age and spectral type dependency
of stellar rotation is the connection to magnetic field
strength and coronal activity.
A low fraction of field stars earlier than M4 are active
stars (e.g. have H$\alpha$ in emission), while the fraction
sharply increases with spectral type and finally
culminates at $\sim$M8V, where more than 70\% of all field
stars show signs of activity (see Fig 2).
The question remains at which age this activity
relationship is fully established. From our FLAMES
study we have found that among five Coma members
with spectral types of M0V--M2V,
two members exhibit strong H$\alpha$ emission, while only two members are
inactive. This indicates a fraction of about 50\% of active
stars with spectral types of early-M,
still significantly more than in the old field population. All of the Coma
cluster members in our sample, even the highly active
M2V, have rotational velocities of less than 5 km/s.

\section{IC2602: Cluster memberships reappraised}

Our FLAMES sample in IC 2602 (age $\sim35$\,Myrs) was
based on a list of objects from Foster et al. (\cite{1997A&AS..126...81F}, hereafter F97)
and from Scholz \& Eisl\"offel (private communication). By
comparing the radial velocities obtained with FLAMES to
the mean radial velocity of IC 2602 ($+16$\,km/s, Randich
et al., \cite{2001A&A...372..862R}), we can exclude single targets with deviant
radial velocities (see Fig 3). In addition we evaluated the
Lithium EW and the activity status of the targets to
further constrain their membership.
We find that only a minority of the stars given in F97 are
likely to be actual members of IC 2602. This is consistent
with the locus of the stars in a colour--colour
diagram, where most of the targets appear reddened by $A_v\sim2$\,mag,
but actual cluster members are found close to the un-reddend
main sequence.

\begin{figure}[!ht]
\resizebox{0.8\textwidth}{!}{
\includegraphics[]{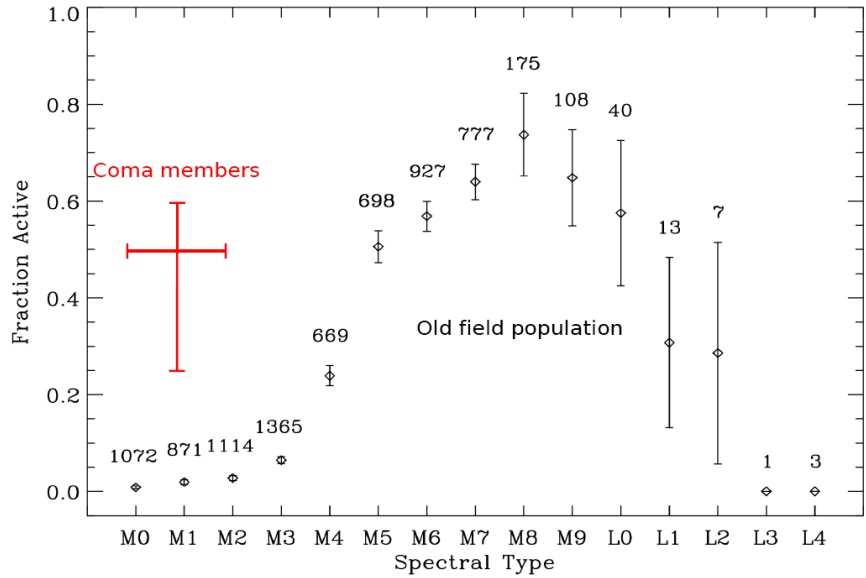}
}
\caption{The fraction of active field stars over spectral types of M0V--L4V 
from a large SDSS sample (figure from West et al. \cite{2004AJ....128..426W}, reproduced
by permission of the AAS).
The numbers above each bin represent the total number of stars
used to compute the fraction. The red datapoint marks the
fraction of active stars found in our FLAMES survey of the Coma
cluster (age $\sim1$\,Gyr).}
\end{figure}

\begin{figure}[!hb]
\resizebox{.8\textwidth}{!}
{\includegraphics[]{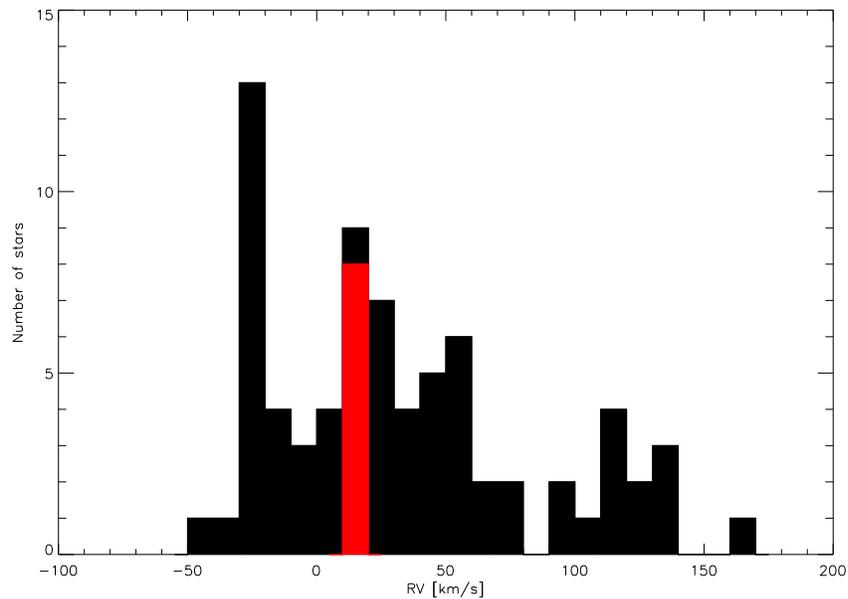}}
  \caption{The radial velocities of all IC 2602 targets in our FLAMES
survey. Only eight out of 75 targets have a radial velocity
consistent with the value expected for IC 2602 ($+16$\,km/s, based
on a sample by Randich et al., \cite{2001A&A...372..862R}). However, only two of these
eight targets show the expected Li(6710) absorption and thus
classify as candidate members.}
\end{figure}

\clearpage

\begin{figure}[!ht]
\resizebox{.8\textwidth}{!}
{\includegraphics[]{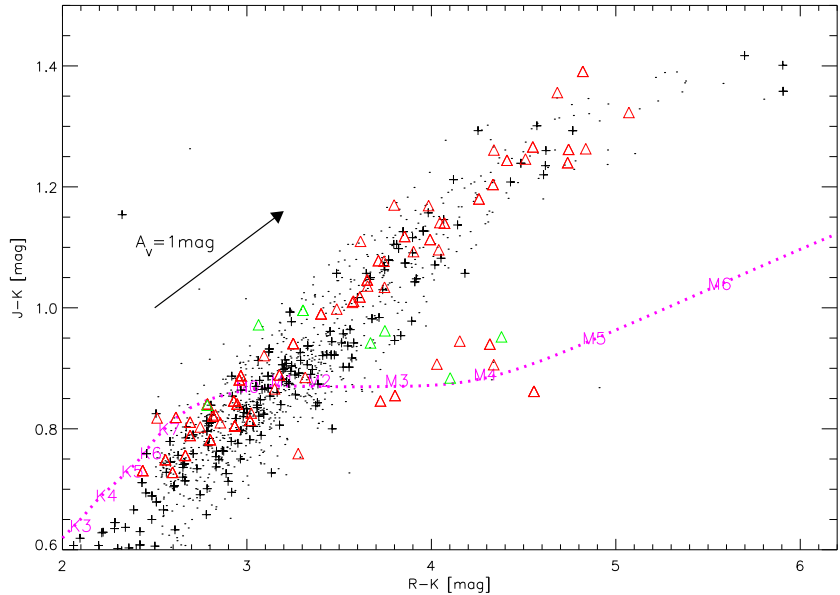}}
  \caption{Colour--colour diagram for stars in the field of the young
cluster IC 2602: all UCAC2 stars (dots), observed UCAC2 stars
(crosses), target from the list of F97 (red triangles), and other
candidate members (green triangles). The main sequence is
represented by a magenta line and a reddening vector for $A_v=1$\,mag
is shown. Note that most of the objects from F97 form a
reddened main-sequence track and are not members of IC 2602.}
\end{figure}

%\begin{figure}
%\resizebox{\textwidth}{!}
%{\includegraphics[bb = -48 120 661 713,clip]{GJ860_poster}}
%  \caption{An example of the data obtained with the ULBCAM, the rich
%stellar field around GJ 860. GJ860 is partially hidden behind the
%gap of the detector mosaic. The total FOV is about 23$\times$23\,
%arcmin. The full depth of H$_\mathrm{limit}\sim19$\,mag is reached 
%in the central 16$\times$16\,arcmin.}
%\end{figure}

%%%%%%%%%%%%%%%%%%%%%%%%%%%%%%%%%%%%%%%%%%%%%%%%
%% BACKMATTER
%%%%%%%%%%%%%%%%%%%%%%%%%%%%%%%%%%%%%%%%%%%%%%%%

\begin{theacknowledgments}
AR and AS acknowledge research funding from the DFG under an Emmy Noether Fellowship (RE 1664/41).
\end{theacknowledgments}

%%%%%%%%%%%%%%%%%%%%%%%%%%%%%%%%%%%%%%%%%%%%%%%%
%% The bibliography can be prepared using the BibTeX program or
%% manually.
%%
%% The code below assumes that BibTeX is used.  If the bibliography is
%% produced without BibTeX comment out the following lines and see the
%% aipguide.pdf for further information.
%%
%% For your convenience a manually coded example is appended
%% after the \end{document}
%%%%%%%%%%%%%%%%%%%%%%%%%%%%%%%%%%%%%%%%%%%%%%%%

%%%%%%%%%%%%%%%%%%%%%%%%%%%%%%%%%%%%%%%%%%%%%%%%
%% You may have to change the BibTeX style below, depending on your
%% setup or preferences.
%%
%%
%% For The AIP proceedings layouts use either
%%%%%%%%%%%%%%%%%%%%%%%%%%%%%%%%%%%%%%%%%%%%

\bibliographystyle{aipproc}   % if natbib is available
%\bibliographystyle{aipprocl} % if natbib is missing

%%%%%%%%%%%%%%%%%%%%%%%%%%%%%%%%%%%%%%%%%%%
%% You probably want to use your own bibtex database here
%%%%%%%%%%%%%%%%%%%%%%%%%%%%%%%%%%%%%%%%%%%
%\bibliography{sample}

%%%%%%%%%%%%%%%%%%%%%%%%%%%%%%%%%%%%%%%%%%%
%% Just a reminder that you may have to run bibtex
%% All of it up to \end{document} can be removed
%% if you don't like the warning.
%%%%%%%%%%%%%%%%%%%%%%%%%%%%%%%%%%%%%%%%%%%
\IfFileExists{\jobname.bbl}{}
 {\typeout{} \typeout{******************************************}
  \typeout{** Please run "bibtex \jobname" to optain} \typeout{** the
  bibliography and then re-run LaTeX} \typeout{** twice to fix the
  references!}  \typeout{******************************************}
  \typeout{} }

\end{document}